\documentstyle{mn}

\input epsf

\begin{document}

\def\Lya{Ly$\alpha\ $}
\def\Lyb{Ly$\beta\ $}
\def\Lyg{Ly$\gamma\ $}
\def\Lyd{Ly$\delta\ $}
\def\Lye{Ly$\epsilon\ $}
\def\LCDM{$\Lambda$CDM\ }
\def\HI{\hbox{H~$\rm \scriptstyle I\ $}}
\def\HII{\hbox{H~$\rm \scriptstyle II\ $}}
\def\DI{\hbox{D~$\rm \scriptstyle I\ $}}
\def\HeI{\hbox{He~$\rm \scriptstyle I\ $}}
\def\HeII{\hbox{He~$\rm \scriptstyle II\ $}}
\def\HeIII{\hbox{He~$\rm \scriptstyle III\ $}}
\def\CIV{\hbox{C~$\rm \scriptstyle IV\ $}}
\def\NHI{N_{\rm HI}}
\def\NHeII{N_{\rm HeII}}
\def\cm2{\,{\rm cm$^{-2}$}\,}
\def\kms{\,{\rm km\,s$^{-1}$}\,}
\def\skm{\,({\rm km\,s$^{-1}$})$^{-1}$\,}
\def\kmsmpc{\,{\rm km\,s$^{-1}$\,Mpc$^{-1}$}\,}
\def\hmpc{\,h^{-1}{\rm \,Mpc}\,}
\def\mpch{\,h{\rm \,Mpc}^{-1}\,}
\def\hkpc{\,h^{-1}{\rm \,kpc}\,}
\def\ev{\,{\rm eV\ }}
\def\kel{\,{\rm K\ }}
\def\intunits{\,{\rm ergs\,s^{-1}\,cm^{-2}\,Hz^{-1}\,sr^{-1}}}
\def\ltsima{$\; \buildrel < \over \sim \;$}
\def\lsim{\lower.5ex\hbox{\ltsima}}
\def\gtsima{$\; \buildrel > \over \sim \;$}
\def\gsim{\lower.5ex\hbox{\gtsima}}
\def\etal{{ et~al.~}}
\def\aj{AJ}
\def\ana{A\&A}
\def\apj{ApJ}
\def\apjs{ApJS}
\def\mn{MNRAS}

\journal{Preprint-03}

\title{Radiative transfer through the Intergalactic Medium}

\author[J. Bolton, A. Meiksin and M. White]{James Bolton${}^{1}$,
Avery Meiksin${}^{2}$, Martin White${}^{3}$ \\
${}^1$Institute of Astronomy, University of Cambridge,
The Observatories, Madingley Road, Cambridge CB3\ 0HA, UK \\
${}^2$Institute for Astronomy, University of Edinburgh,
Blackford Hill, Edinburgh\ EH9\ 3HJ, UK \\
${}^3$Departments of Astronomy and Physics, University of California,
Berkeley, CA 94720, USA}

%\pubyear{2003}

\maketitle

\begin{abstract}
We use a probabilistic method to compute the propagation of an
ionization front corresponding to the re-ionization of the
intergalactic medium in a $\Lambda$CDM cosmology, including both
hydrogen and helium. The effects of radiative transfer
substantially boost the temperature of the ionized gas over the case
of uniform re-ionization. The resulting temperature-density relation
of the ionized gas is both non-monotonic and multiple-valued,
reflecting the non-local character of radiative transfer and
suggesting that a single polytropic relation between local gas density
and temperatue is a poor description of the thermodynamic state of baryons
in the post-reionization universe.
\end{abstract}

%%%%%%%%%%%%%%%%%%%%%%%%%%%%%%%%%%%%%%%%%%%%%%%%%%%%%%%%%%%%%%%%%%%%%%%%%%%%%%%

\section{Introduction} \label{sec:introduction}

Hydrodynamical simulations of structure formation in the universe have
led to fundamental insights into the structure and evolution of the
Intergalactic Medium (IGM) (Cen \etal 1994; Zhang, Anninos \& Norman
1995; Hernquist \etal 1996; Zhang \etal 1997; Bond \& Wadsley 1997;
Theuns, Leonard \& Efstathiou 1998). Comparisons with the \Lya forest,
as measured in high redshift Quasi-Stellar Object (QSO) spectra, show
that the simulations broadly reproduce the statistical properties of
the IGM. Precision comparisons with the highest spectral resolution
measurements recover the cumulative flux distributions and \HI column
density distributions to an accuracy of a few percent (Meiksin, Bryan
\& Machacek 2001). More problematic are the predicted widths of the
absorption features, which appear to require additional sources of
broadening, perhaps late \HeII re-ionization, to match the measured
widths (Theuns \etal 1999; Bryan \& Machacek 2000; Meiksin \etal 2001).

Many of the properties of the IGM may be attributed to the
gravitational instability of the dark matter alone, as the baryon
density fluctuations closely follow those of the dark matter (Zhang
\etal 1998). This has led to the development of pseudo-hydrodynamic
simulations based on pure gravity in which the baryon fluctuations are
assumed to exactly follow the dark matter and the temperature to be
derived from an effective equation of state (Petitjean, M\"ucket \&
Kates 1995; Croft \etal 1998; Gnedin \& Hui 1998; Meiksin \& White
2001).
In recent years, such simulations have been increasingly relied on for
predicting the flux power spectrum of the \Lya forest (Meiksin \& White 2001;
Zaldarriaga, Hui \& Tegmark 2001; Croft \etal 2002; Meiksin \& White 2003b;
Seljak, McDonald \& Makarov 2003).

The simulations have generally been done assuming sudden early
homogeneous \HI and \HeII re-ionization. While tentative steps have
been taken to introduce radiative transfer into the simulations using
approximate schemes (Abel, Norman \& Madau 1999; Gnedin \& Abel 2001;
Nakamoto, Umemura \& Susa 2001; Ciardi, Stoehr \& White 2003), these
simulations have emphasized the propagation of \HI ionization fronts
and the predicted mean optical depths. Except for the simulations of
Gnedin \& Abel, they have not solved fully self-consistently for the
combined gas dynamics and radiative transfer.

In this Letter, we extend the photon-conserving scheme of Abel \etal to
include the treatment of helium and explore the thermal effects of
radiative transfer in the inhomogeneous medium predicted in a $\Lambda$CDM
cosmology.

\section{Radiative transfer}

\subsection{Simulation data}
\begin{figure}
\begin{center}
\leavevmode 
\epsfxsize=2.1in
\epsfysize=2.1in
\epsfbox{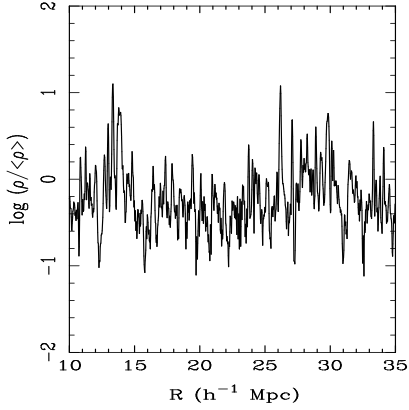}
\end{center}
\caption{The baryon density fluctuations at $z=6$ against comoving distance
from the ionizing source for the line of sight considered in our
radiative transfer simulations.}
\label{fig:baryon}
\end{figure}

We use the data from a previously run $\Lambda$CDM simulation used to
investigate effects of radiative transfer
on the \Lya forest (Meiksin \& White 2003b). The simulation was
run using a pure Particle Mesh (PM) dark matter code, and it was
assumed the gas and dark matter have the same spatial distribution.
A description of the parallel PM code is given in Meiksin \& White (2003a).
The parameters used for the simulation are $\Omega_{\rm M}=0.30$,
$\Omega_\Lambda=0.70$, $\Omega_b=0.045$, $h=H_0/100$\kmsmpc$=0.70$, and
slope of the primordial density perturbation power spectrum $n=1.05$.
The model is consistent with existing large-scale structure, \Lya
forest flux distribution, cluster abundance and WMAP constraints
(Meiksin \& White 2003b). The simulation was run using $512^3$
particles and a $1024^3$ force mesh, in a cubic box with (comoving)
side length $25\,h^{-1}$Mpc, adequate for obtaining converged estimates
of the \Lya pixel flux distribution and flux power spectrum (Meiksin \&
White 2003a,b). A typical line-of-sight density run
at $z=6$, used for the radiative transfer calculations here,
is shown in Fig.~\ref{fig:baryon}.

\subsection{Equations of radiative transfer}
\label{sec:RT}

To consider the effect of radiative transfer when modelling the
propagation of ionization fronts (I-fronts) into the IGM, we extend
the photon-conserving algorithm of Abel \etal to include helium.
The advantage of this scheme is that energy is conserved
independently of numerical resolution, ensuring that I-fronts
propagate at the correct speeds in our simulations. In practice, this
means that larger step sizes may be taken on the simulation space grid
without the associated loss of accuracy which would occur when solving
the radiative transfer equation through direct numerical integration.

The probabilities for the absorption of an ionizing photon
by \HI, \HeI and \HeII, respectively, are

\begin{eqnarray}
P_{\rm abs}^{\rm HI} &=& p_{\rm HI}q_{\rm HeI}q_{\rm HeII}
\left[1-\exp\left(-\tau_{\nu}^{\rm tot}\right)\right]/D, \\
P_{\rm abs}^{\rm HeI} &=& q_{\rm HI}p_{\rm HeI}q_{\rm HeII}
\left[1-\exp\left(-\tau_{\nu}^{\rm tot}\right)\right]/D, \\
P_{\rm abs}^{\rm HeII} &=& q_{\rm HI}q_{\rm HeI}p_{\rm HeII}
\left[1-\exp\left(-\tau_{\nu}^{\rm tot}\right)\right]/D.
\end{eqnarray}

\noindent Here we have defined the auxiliary absorption and
transmission probabilities $p_i=1-\exp(-\tau_{\nu}^i)$,
$q_i=\exp(-\tau_{\nu}^i)$, $i$ denoting the species being referred to,
$\tau_{\nu}^{\rm tot}=\tau_{\nu}^{\rm HI}+\tau_{\nu}^{\rm HeI}+
\tau_{\nu}^{\rm HeII}$, where $\tau_{\nu}^{i}$ is the optical depth
for a given species, and $D=p_{\rm HI}q_{\rm HeI}q_{\rm HeII}+q_{\rm
HI} p_{\rm HeI}q_{\rm HeII}+q_{\rm HI}q_{\rm HeI}p_{\rm HeII}$. These
probabilities can be used to calculate the ionization rate for a given
species per unit volume, $n_i\Gamma_i$, as follows. If in a time
$\delta t$, $\delta t \dot N_{\nu}^{l-1}$ photons enter grid zone $l$
from zone $l-1$, then the number of photons that will be absorbed in
zone $l$ by species $i$ is $\delta t \dot N_{\nu}^{l-1}P_{\rm abs}^i$,
where $P_{\rm abs}^i$ is the absorption probability for species $i$
within zone $l$. The ionization rate of species $i$ in zone $l$ of
volume $V^l$ is then
\begin{equation}
n^l_i\Gamma^l_i = \frac{1}{V^l}\sum_g \dot N_{\nu_g}P_{\rm abs}^i(\nu_g),
\label{eq:Gamma}
\end{equation}
where we have divided the frequencies into 100 discrete groups $g$ evenly
spaced between $\nu^{\rm HI}_L$ and $10\nu^{\rm HI}_L$. (While we found
this spacing to be adequate, we have made no attempt to optimize it.)
These are used in the following set of coupled equations to solve
for the positions of the three I-fronts over time:

\begin{equation}
 \frac{dn_{\rm HII}}{dt}\ =n_{\rm HI}\Gamma_{\rm HI}-n_{e}n_{\rm HII}\alpha_{\rm HI}(T)-3\frac{\dot{a}}{a}n_{\rm HII},
\label{eq:HII}
\end{equation}

\begin{eqnarray}
 \frac{dn_{\rm HeII}}{dt}\ &=& n_{\rm HeI}\Gamma_{\rm HeI}+n_{e}n_{\rm HeIII}\alpha_{\rm HeII}(T)-n_{\rm HeII}\Gamma_{\rm HeII} \nonumber\\
&-&n_{e}n_{\rm HeII}\alpha_{\rm HeI}(T)-3\frac{\dot{a}}{a}n_{\rm HeII},
\label{eq:HeII}
\end{eqnarray}

\begin{equation}
 \frac{dn_{\rm HeIII}}{dt}=n_{\rm HeII}\Gamma_{\rm HeII}-n_{e}n_{\rm HeIII}\alpha_{\rm HeII}(T)-3\frac{\dot{a}}{a}n_{\rm HeIII},
\label{eq:HeIII}
\end{equation}  

\noindent where $n_i$ denotes number density, $\Gamma_i$ $({\rm
s^{-1}})$ is the photoionization rate per atom, $\alpha_i(T)$ is the
total radiative recombination coefficient, and $a$ is the cosmological
expansion factor.

The time evolution of the gas temperature $T$ is given by:

\begin{equation}
\frac{dT}{dt}\ = \frac{2(G-L)}{3kn}\ + \frac{T}{n}\ \frac{dn_{e}}{dt}\
-3\frac{\dot{a}}{a}\left(\frac{2}{3}T + \frac{n_{e}}{n}T
\right).
\label{eq:thermal}
\end{equation}

\noindent where $n=n_{\rm HI}+n_{\rm HeI}+n_{\rm HII}+n_{\rm
HeII}+n_{\rm HeIII}+n_{e}$, $n_{e}=n_{\rm HII}+n_{\rm HeII}+2n_{\rm
HeIII}$, $k$ is the Boltzmann constant, and $G$ and $L$ $({\rm
J\,m^{-3}\,s^{-1}})$ are the atomic heating and cooling rates,
respectively. The last term is the adiabatic cooling term resulting
from cosmological expansion, which will dominate the thermal effects
of gas motions at the densities we consider.  While computing the
photoionization, we assume the gas over-density (not the density)
stays frozen, which is a good approximation on the scales relevant to
the \Lya forest (Zhang \etal 1998). As in similar previous studies, we
do not include the effects of adiabatic compression heating as this
will only affect the temperature for the relatively rare virialised
halos with circular velocities exceeding $\sim20$\kms (Meiksin 1994;
Meiksin \& White 2003b).

The heating rate $G^l$ in cell $l$ is due to the photoionization of \HI, \HeI
and \HeII according to $G^l=G^l_{\rm HI} + G^l_{\rm HeI} + G^l_{\rm HeII}$,
where for each species $i$, $G^l_i$ is evaluated in a similar manner to the
ionization rate per unit volume as given in Eq.~(\ref{eq:Gamma}),

\begin{equation}
n^l_i G^l_i = \frac{1}{V^l}\sum_g (h\nu_g - \chi_i)\dot N_{\nu_g}
P_{\rm abs}^i(\nu_g),
\label{eq:G}
\end{equation}
where $\chi_i$ is the ionization potential of species i.
The atomic cooling rate $L$ includes radiative recombination cooling to
\HI, \HeI and \HeII, electron excitation of \HI, and Compton cooling off
the Cosmic Microwave Background photons. The radiative recombination and
cooling rates are taken from Meiksin (1994).

Eqs.~(\ref{eq:HII})--(\ref{eq:thermal}) are solved using explicit
forward Euler integration. Although an implicit numerical scheme such
as backward Euler integration results in a more stable solution at low
numerical resolution (Anninos \etal, 1997), we find that to obtain the
required numerical accuracy as well as to maintain stability, both
methods require similar numerical resolution. For the sake of speed
and simplicity, the forward method is preferred. The timestep is
restricted to being no more than several times greater that the
hydrogen ionization timescale, $\Gamma_{\rm HI}^{-1}$, for numerical
accuracy.

We take the frequency specific luminosity of the ionizing source in
our simulations to be described by a power law spectrum with index
$\alpha=1.5$, and a luminosity of $L_\nu^Q=10^{23}\, {\rm W\,
Hz^{-1}}$ at the Lyman edge. This spectrum is typical of that of QSOs,
which are probable candidate sources for contributing to the
re-ionization of the IGM. The assumed mass fraction of helium in the
IGM is $Y\simeq0.235$.

We tested the photon-conserving algorithm on the photoionization of
gas with uniform density around a point source. The resulting
solutions for the ionization and temperature profiles were accurate to
better than 10 per cent for optical depths at the Lyman edge of up to
$\Delta\tau_{\nu}\simeq20$ per cell on the space grid. This results in
a significant reduction in the computational resources compared with a
radiative transfer scheme in which the ionization rates are computed
by direct numerical integration over the intensity
($L_\nu^Qe^{-\tau_\nu}/[4\pi r]^2$) and photoionization cross-sections
(eg, Madau, Meiksin \& Rees 1997). Typically, we find that to obtain
an accuracy comparable to the photon-conserving scheme at
$\Delta\tau_{\nu}^{\rm HI}\simeq20$, a direct integration algorithm
requires $\Delta\tau_{\nu}^{\rm HI}\simeq1/4$ in each cell. For the
simulation results presented in this Letter, the original PM grid of
1024 separate cells was used over a line of sight 25$h^{-1}$ comoving
Mpc in length. This provided accurate spatial resolution, as at most
$\Delta\tau_{\nu}^{\rm HI}\simeq6.8$ in each cell at the average
baryonic density assumed for the neutral IGM at $z=6$.

\section{Results}
\label{sec:results}

We set up a problem of a QSO source of luminosity $L_\nu^Q$ at $z=6$
placed at a comoving distance of 10$h^{-1}$~Mpc from the left edge of
the density run shown in Fig.~\ref{fig:baryon}, and assume the gas
surrounding the QSO up to that point has already been ionized. Placing
the source at this distance avoids having to impose the restriction
that the I-fronts propagate no faster than the speed of light. We
indicate the displaced position of the source in the figures by
starting the (comoving) spatial axes at $R=10h^{-1}$~Mpc.

Two cases were compared:\ computing the ionization and
temperature profiles by incorporating radiative transfer as discussed
in section \ref{sec:RT} (hereafter the RT simulation), and a second
case neglecting radiative transfer, assuming instead an instantaneous
uniform ionization rate across the whole line of sight (hereafter the
nRT simulation), as is usually done in simulations of the \Lya
forest. Our intention is to compare the final gas temperatures to
determine how large an effect including radiative transfer may have.

\begin{figure}
\begin{center}
\leavevmode 
\epsfxsize=2.1in
\epsfysize=2.1in
\epsfbox{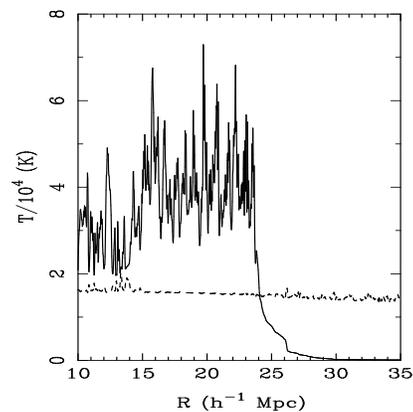}
\end{center}
\caption{Comparison of the IGM temperature at $z=5$ against comoving
distance from the ionizing source. The solid line is computed using
radiative transfer, while the dashed line result assumes a uniform
ionization field through the entire line of sight.}
\label{fig:tr5}
\end{figure}

\begin{figure}
\begin{center}
\leavevmode 
\epsfxsize=2.1in
\epsfysize=2.1in
\epsfbox{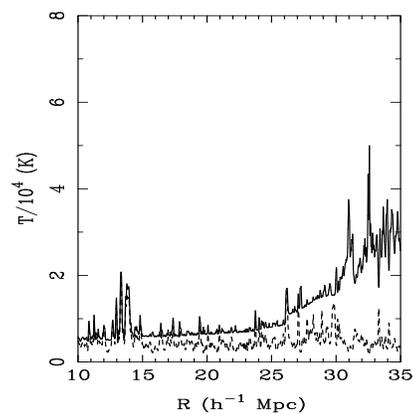}
\end{center}
\caption{Comparison of the IGM temperature at $z=3$ against comoving
distance from the ionizing source. The line types are as in
Fig.~\ref{fig:tr5}.}
\label{fig:tr3}
\end{figure}

Fig.~\ref{fig:tr5} shows a comparison between the IGM temperatures at
$z=5$ computed with and without radiative transfer.  The inclusion of
radiative transfer results in a significant boost to the temperature
of the ionized IGM, a point illustrated by Abel \& Haehnelt (1999) in
the context of a smooth IGM. The positions of the \HII and \HeII
I-fronts at $z=5$ in the RT simulation are 24$h^{-1}$ comoving Mpc
from the source, while the \HeIII front lags behind at about
19$h^{-1}$ comoving Mpc; their speeds of propagation are limited by
the reduction in the ionization rate due to the increasing optical
depth through the box. Consequently, the heating of the IGM, which we
find to be dominated by the photoionization of neutral hydrogen at the
\HII I-front, is also constrained to lie behind the \HII I-front. All
gas beyond 24$h^{-1}$ Mpc is still much cooler. In contrast, the
temperature computed by the nRT simulation is much lower and the
heating extends across the whole line of sight.  The temperature is
also much less sensitive to variations in the gas density for the nRT
simulation. Fig.~\ref{fig:tr3} compares the results
obtained at $z=3$. By this time, both the \HII and \HeII I-fronts have
reached 35$h^{-1}$ Mpc from the source.  The material which has been
ionized early on in the RT simulation is now beginning to cool to the
same temperature as that computed by the nRT simulation, although the
temperature differences remain large.  In particular, the gas that was
photoionized at $z=5$ ($R<24\hmpc$), remains nearly twice as hot as
the uniformly ionized gas by $z=3$.  By contrast, the regions which
were ionized more recently are much hotter than for the nRT
simulation. We find that only by $z=1$, once the gas has been in an
ionized state for a sufficiently long period of time, do the
temperatures predicted in the two scenarios converge.

\begin{figure}
\begin{center}
\leavevmode 
\epsfxsize=2.1in
\epsfysize=2.1in
\epsfbox{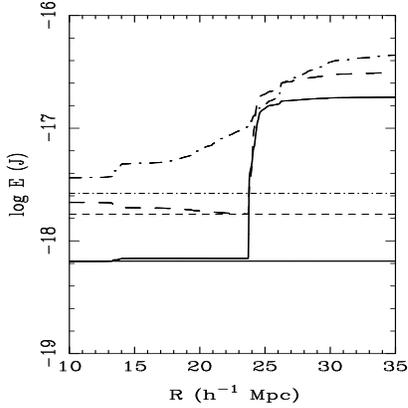}
\end{center}
\caption{Plot of the energy per ionization for HI (solid line) HeI
(dashed line) and HeII (dot-dashed line) at $z=5$. The bold lines
correspond to the results computed using radiative transfer, while the
lighter (flat) lines correspond to the uniform ionization field.}
\label{fig:energy}
\end{figure}

The reason for the temperature boost in the RT simulation is made
clear by considering the energy per ionization ($E=G/\Gamma$) at the
I-front. In Fig.~\ref{fig:energy}, we show that the average energy
of a photon absorbed by either \HI, \HeI or \HeII increases at the
relevant I-front, as proportionally more of the lower energy photons
have been absorbed already -- a consequence of including radiative
transfer. This increase to the heating energy per ionization is
responsible for the temperature boost. The assumption of an
instantaneous ionization field across the line of sight results in an
underestimate of the IGM temperature at high redshift.

\begin{figure}
\begin{center}
\leavevmode 
\epsfxsize=2.1in
\epsfysize=2.1in
\epsfbox{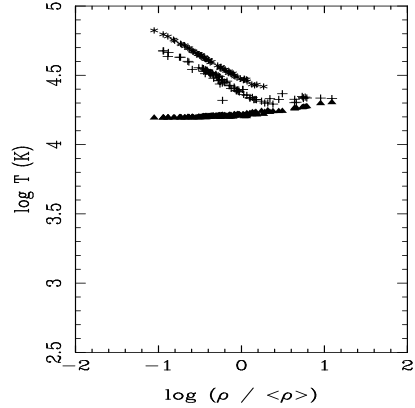}
\end{center}
\caption{Comparison of the IGM temperature-density relation at $z=5$ for
regions with fully ionized helium
($n_{HeIII}/n_{He}>0.9$). The triangles correspond to the
uniform ionization case. The data points for the
results incorporating radiative transfer are split into two sets:\
$+$ corresponds to gas in the range $10h^{-1}<R<14h^{-1}$, for which
the mean over-density $\langle\rho/\bar\rho\rangle=1.452$, (where
$\bar\rho$ is the global mean gas density); and
$\ast$ designates gas in the range $14h^{-1}<R<23.6h^{-1}$, with mean
over-density 0.519.}
\label{fig:td5}
\end{figure}

\begin{figure}
\begin{center}
\leavevmode 
\epsfxsize=2.1in
\epsfysize=2.1in
\epsfbox{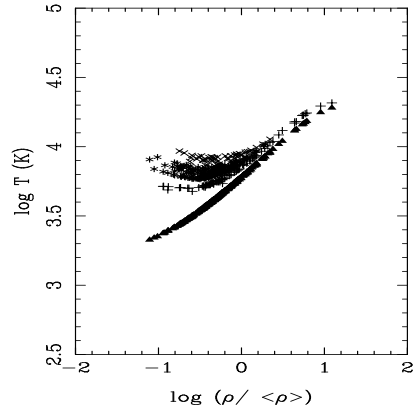}
\end{center}
\caption{Comparison of the IGM temperature-density relation at $z=3$. The
symbols are as in Fig.~\ref{fig:td5}. We also add a third point
$\times$ designating gas in the range $23.6h^{-1}<R<25.8h^{-1}$, with mean
over-density 0.671.}
\label{fig:td3}
\end{figure}

\begin{figure}
\begin{center}
\leavevmode 
\epsfxsize=2.1in
\epsfysize=2.1in
\epsfbox{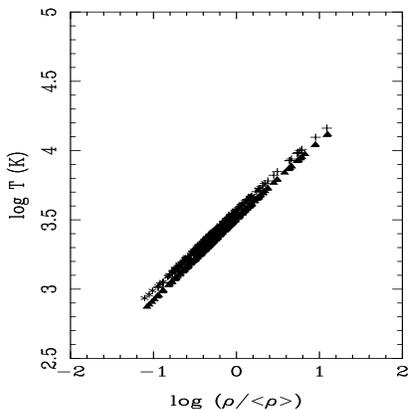}
\end{center}
\caption{Comparison of the IGM temperature-density relation at $z=1$. The
symbols are as in Figs.~\ref{fig:td5} and \ref{fig:td3}.}
\label{fig:td1}
\end{figure}

A comparison of the IGM temperature-density relation computed by the
two different simulations is made in Figs.~\ref{fig:td5}--\ref{fig:td1}.
Only data points where \HeII has been
90 per cent ionized are plotted, so that gas which is neutral or in the
process of being ionized is not considered. In Fig.~\ref{fig:td5},
the underdense regions in the line of sight (Fig.~\ref{fig:baryon}),
are initially heated to a higher temperature in the RT simulation.  A
non-monotonic temperature-density relation results. The relation
actually splits into two trends. To understand the origin of the
split, we divide the data points into two distinct sets corresponding
to two regions in which the average density fluctuations are slightly
higher or lower than the overall average in the line of sight (see
caption to Fig.~\ref{fig:td5}). The gas in the denser region, which
is closer to the source and thus is ionized sooner, shows the same
overall trend as the gas in the less dense region but is slightly
cooler at a fixed over-density. In contrast, in the nRT simulation the
temperature-density relation is both monotonic and essentially
single-valued. By $z=3$ (Fig.~\ref{fig:td3}), the ionized gas has
cooled adiabatically, with the overdense regions cooling at a slower
rate than the underdense regions, as the denser regions are better
able to maintain thermal balance because of the shorter cooling (and
photoionization heating) times. The IGM temperatures computed by the
two simulations are now beginning to converge at high density, but at
low density there is still a significant difference between the
results.  A third set of data points is also evident, corresponding to
the latest locally underdense region to be ionized.
 
Finally, by $z=1$, the gas temperatures computed by the RT
and nRT simulations have nearly converged. The gradient of the line in
the temperature-density plane indicates that adiabatic cooling due to
expansion is largely responsible for cooling the gas. We find the
entropy is now nearly constant throughout the line of sight. The
striations in the RT data points have also disappeared. By $z=1$, the
results from the two simulations are in good agreement.

\section{Conclusions}

We extend the probabilistic method of radiative transfer described for
hydrogen by Abel \etal (1999) to include helium and tested the method
on the propagation of an ionization front through a uniform medium. We
compare the method against direct numerical integration of the exact
time-steady radiative transfer equation. We confirm their conclusion
that the probabilistic method accurately reproduces the position of
the ionization fronts even for large optical depths at the
photoelectric edges. Specifically, we recover the numerically
integrated converged solutions for the ionization fractions and
temperature to an accuracy of 10 per cent for an incremental optical depth
per grid zone at the \HI photoelectric edge as high as 20, while
direct numerical integration of the radiative transfer equations
requires an incremental optical depth less than 1/4 to obtain
comparable accuracy.

We use the probabilistic method to compute the propagation of an
I-front generated by a QSO source through the IGM, using the density
distribution drawn from a numerical simulation in a $\Lambda$CDM
universe. Including the effects of radiative transfer results in a
substantial boost in the post-ionized gas temperature compared with
the case of sudden uniform ionization with no radiative transfer. The
boost is due chiefly to the increased energy per photoionization
within the ionization fronts, a consequence of the higher probability
for high energy photons to pass through the gas compared with lower
energy photons.

A consequence of the radiative transfer is to alter the dependence of
temperature on density from the approximately polytropic relation
found for moderate to low densities assuming uniform photoionization
(Meiksin 1994; Gnedin \& Hui 1998) to a generally non-monotonic trend.
While the differences between the temperatures for over-densities
above $\sim5$ are small (less than 25 per cent), at lower densities the two
temperatures converge only slowly with time, nearly recovering the
polytropic relation only by $z\approx1$.

At $z>1$, the relation between temperature and density is not
single-valued:\ multiple temperatures may result for different gas
parcels of the same over-density. Although this is in part a
consequence of the delay in time for gas at different distances from
the source to be photoionized as the ionization front propagates, an
additional contributing factor is the environment of the gas parcel:\
if the ionization front has passed through a region of dense gas,
sufficient to drive the optical depth at the \HI photoelectric edge to
near unity, before reaching the parcel, the radiation field will
deliver a higher energy per photoionization to the parcel. Thus the
determination of the gas temperature is strongly non-local due to the
dependence of the heating rate on the optical depth to photoionizing
photons between the gas parcel and the source of ionizing radiation.

Although the inclusion of radiative transfer results in a substantial
and persistent boost in gas temperature, it is insufficient in itself
to account for the broader measured \Lya absorption systems than
predicted by numerical simulations assuming sudden uniform
photoionization. Meiksin, Bryan \& Machacek (2001) find that a boost
of $\Delta T\approx1.7\times10^4\kel$ is required at $z\lsim3.5$,
larger than the temperatures we find here. Although it may be possible
to obtain a larger boost for an alternative re-ionization scenario,
such a large boost may still require invoking late \HeII
re-ionization. Our principal conclusion here is that for numerical
simulations to accurately predict the temperature of the IGM, they
must include the effects of radiative transfer. We are currently
extending our methods to incorporate radiative transfer with ray
tracing (eg, Abel \& Wandelt 2002), into fully self-consistent
simulations to explore this and related issues, including alternative
sources of re-ionization like low luminosity AGNs and stars.

\bigskip
A.M. thanks the University of Edinburgh Development Trust for its
financial support.
%%%%%%%%%%%%%%%%%%%%%%%%%%%%%%%%%


\begin{thebibliography}{}

\bibitem[1999]{ah99}
        Abel T., Haehnelt M.~G., 1999, \apj, 520, L13

\bibitem[1999]{anm99}
        Abel T., Norman M.~L., Madau P., 1999, \apj, 523, 66

\bibitem[2002]{aw02}
        Abel T., Wandelt B.~D., 2002, \mn, 330, L53

\bibitem[1997]{ann97}
	Anninos P., Zhang Y., Abel T., Norman M., 1997, NewA, 2, 209

\bibitem[]{bon97}
        Bond J.~R., Wadsley J.~W., 1997, in Petitjean P., Charlot S., eds,
        Structure and Evolution of the Intergalactic Medium from QSO
        Absorption Line Systems. Editions Fronti\`eres, Paris, p. 143

\bibitem[]{bm00}
        Bryan G.~L., Machacek M.~E., 2000, \apj, 534, 57

\bibitem[]{cen94}
        Cen R., Miralda-Escud\'e J., Ostriker J.~P., Rauch M.,
        1994, \apj, 437, L9

\bibitem[]{csw02}
        Ciardi B., Stoehr F., White S.~D.~M., 2003, \mn, 343, 1101

\bibitem[1998]{cwkh98}
        Croft R.A.C., Weinberg D.~H., Katz N., Hernquist L., 1998, \apj,
        495, 44

\bibitem[]{cro02}
        Croft R.~A.~C., \etal, 2002, \apj, 581, 20

\bibitem[]{ga01}
        Gnedin N.~Y., Abel T., 2001, NewA, 6, 437

\bibitem[]{gh98}
        Gnedin N.~Y., Hui L., 1998, \mn, 296, 44

\bibitem[]{her96}
        Hernquist L., Katz N., Weinberg D., Miralda-Escud\'e J.,
        1996, \apj, 457, L51

\bibitem[]{mmr97}
        Madau P., Meiksin A., Rees M.~J., 1997, \apj, 475, 429

\bibitem[]{mei94}
        Meiksin A., 1994, \apj, 431, 109

\bibitem[]{mbm01}
        Meiksin A., Bryan G.~L., Machacek M.~E., 2001, \mn, 327, 296

\bibitem[2001]{MeiWhi01}
        Meiksin A., White M., 2001, \mn, 324, 141

\bibitem[2003]{PaperI}
        Meiksin A., White M., 2003a, \mn, 342, 1205

\bibitem[2003]{PaperII}
        Meiksin A., White M., 2003b, \mn~(astro-ph/0307289)

\bibitem[2001]{nus01}
        Nakamoto T., Umemura M., Susa H., 2001, \mn, 321, 593

\bibitem[1993]{Pee93}
	Peebles, P.J.E, Principles of Physical Cosmology, 1993, Princeton University Press

\bibitem[1995]{pmk85}
        Petitjean P., M\"ucket J.~P., Kates R.~E., 1995, \ana, 295, L9

\bibitem[2003]{smm03}
        Seljak U., McDonald P., \& Makarov A., 2003, \mn, 342, L79

\bibitem[1998]{the98}
        Theuns T., Leonard A., Efstathiou G., 1998, \mn, 297, L49

\bibitem[1999]{the99}
        Theuns T., Leonard A., Schaye J., Efstathiou G., 1999, \mn, 303, L58

\bibitem[2001]{zht01}
        Zaldarriaga M., Hui L., Tegmark M., 2001, \apj, 557, 519

\bibitem[]{zha95}
        Zhang Y., Anninos P., Norman M.~L., 1995, \apj, 453, L57

\bibitem[]{zman97}
        Zhang Y., Anninos P., Norman M.~L., Meiksin A.,
        1997, \apj, 485, 496

\bibitem[]{zman98}
        Zhang Y., Meiksin A., Anninos P., Norman M.~L.,
        1998, \apj, 495, 63

\end{thebibliography}
\end{document}